&pdflatex

\documentclass[12pt,prd,aps,amssymb,amsmath,tightenlines,showpacs]{revtex4}
\usepackage{graphicx}

\def\half{\textstyle{\frac{1}{2}}}
\def\cPT{\mathcal PT}
\def\cS{\mathcal S}

\begin{document}

\title{Calculation of low-lying energy levels in quantum mechanics}

\author{Carl~M.~Bender${}^1$}
\email{cmb@wustl.edu}

\author{Hugh~F.~Jones${}^2$}
\email{h.f.jones@imperial.ac.uk}

\affiliation{${}^1$Department of Physics, Washington University, St. Louis, MO
63130, USA \\ ${}^2$Blackett Laboratory, Imperial College, London SW7 2AZ, UK}

\begin{abstract}
This paper proposes a very simple perturbative technique to calculate the
low-lying eigenvalues and eigenstates of a parity-symmetric quantum-mechanical
potential. The technique is to solve the time-independent Schr\"odinger
eigenvalue problem as a perturbation series in which the perturbation parameter
is the energy itself. Unlike nearly all perturbation series for physical
problems, for the ground state this perturbation expansion is {\it convergent}
and, even though the ground-state energy is in general not small compared with
1, the perturbative results are numerically accurate. The perturbation series
is divergent for higher energy levels but can be easily evaluated by using
methods such as Pad\'e summation.
\end{abstract}

\maketitle

\section{Introduction}
\label{s1}

This paper presents a very simple idea for calculating the low-lying energy
levels and associated eigenfunctions in quantum mechanics. We begin with the
modest goal of calculating just the ground-state energy $E^{(0)}$ of a
parity-symmetric potential $V(x)$ by treating $E^{(0)}$ as small and using
$E^{(0)}$ itself as a perturbation parameter. Of course, the ground-state energy
is typically not small compared with $1$, and thus one might expect that a
perturbation expansion in powers of $E^{(0)}$ would be so inaccurate as to be
useless. However, for a broad class of potentials the calculation of $E^{(0)}$
turns out to be surprisingly accurate. Moreover, the approximants to $E^{(0)}$
emerge in a form that is ideally suited to Shanks extrapolation, and thus the
numerical accuracy of the calculation can be further improved. Even more
surprising, if the perturbation series for the ground-state energy is converted
to Pad\'e form, the poles and zeros of the Pad\'e approximants converge to the
odd-parity and even-parity energy levels, respectively, and thus the method
proposed here can be used to calculate all of the low-lying energy levels of
$V(x)$.

The specific problem treated here is that of solving the time-independent
Schr\"odinger eigenvalue problem
\begin{equation}
-\psi''(x)+V(x)\psi(x)=E\psi(x),
\label{e01}
\end{equation}
where $\psi(x)$ vanishes as $x\to\pm\infty$. For simplicity, we assume that the
potential $V(x)$ is an {\it even} function of $x$, and to begin we limit our
attention to the {\it ground-state} eigenfunction $\psi(x)$, which is an even
and nodeless function of $x$. We treat the energy $E$ as being small and seek a
perturbation expansion for $\psi(x)$ as a power series in $E$:
\begin{equation}
\psi(x)=\sum_{k=0}^\infty E^k\psi_k(x).
\label{e02}
\end{equation}
We will see that for the ground-state energy $E=E^{(0)}$ this series is
convergent and also that the radius of convergence of the series is the energy
$E^{(1)}$ of the first excited state.

We substitute $\psi(x)$ in (\ref{e02}) into (\ref{e01}) and collect powers of
$E$. The result is the sequence of differential equations
\begin{equation}
\psi_0''(x)=V(x)\psi_0(x)
\label{e03}
\end{equation}
and
\begin{equation}
-\psi_k''(x)+V(x)\psi_k(x)=\psi_{k-1}(x)\quad(k\geq1).
\label{e04}
\end{equation}
Of course, we cannot require that $\psi_0(x)$ satisfy the homogeneous boundary
conditions $\psi_0(\pm\infty)=0$ because $0$ is not an eigenvalue. Instead, we
work on the half-line $0\leq x<\infty$ and impose the {\it inhomogeneous}
boundary conditions $\psi_0(0)=1$ and $\psi_0(+\infty)=0$. The solution $\psi_0
(x)$ to (\ref{e03}) subject to these boundary conditions is unique.

We solve (\ref{e04}) for $\psi_k(x)$ by using the method of reduction of order.
We substitute 
\begin{equation}
\psi_k(x)=\psi_0(x)\phi_k(x),
\label{e05}
\end{equation}
where $\phi_0(x)=1$, and get
\begin{equation}
-\psi_0(x)\phi_k''(x)-2\psi'_0(x)\phi_k'(x)=\psi_0(x)\phi_{k-1}(x).
\label{e06}
\end{equation}
Multiplying by $\psi_0(x)$, the integrating factor for this equation, gives
$$-[\psi_0^2(x)\phi_k'(x)]'=\psi_0^2(x)\phi_{k-1}(x).$$
Using $\psi_0(\infty)=0$ we obtain
\begin{equation}
\phi_k'(x)=\frac{1}{\psi_0^2(x)}\int_x^\infty ds\,\psi_0^2(s)\phi_{k-1}(s).
\label{e07}
\end{equation}

Without loss of generality, we impose the boundary condition $\phi_k(0)=0$ for
$k\geq1$ and integrate (\ref{e07}) again to obtain
\begin{equation}
\phi_k(x)=\int_0^x\frac{ds}{\psi_0^2(s)}\int_s^\infty dt\,\psi_0^2(t)
\phi_{k-1}(t).
\label{e08}
\end{equation}
This equation may be iterated repeatedly to find the expression for $\phi_k(x)$
as a $2k$-fold integral over $\psi_0^2(x)$. The expression for $\phi_k'(x)$ is a
$(2k-1)$-fold integral.

The boundary condition that $\phi_k(0)=0$ is naturally imposed at $x=0$ because
we are assuming that the potential $V(x)$ is symmetric under parity reflection.
An immediate consequence of this requirement seems to be that $\psi(x)$ is
normalized so that $\psi(0)=\psi_0(0)=1$. However, we emphasize that this
conclusion is only valid so long as the perturbation series (\ref{e02})
converges; if $|E|$ exceeds the radius of convergence of the series (\ref{e02}),
we can no longer assume that $\psi(0)=1$. Indeed, when $E$ is an odd-parity
eigenvalue of $V(x)$, $\psi(0)=0$.

To summarize, the explicit expression for $\psi(x)$ in terms of the $\phi_k(x)$
is
\begin{equation}
\psi(x)=\psi_0(x)\left[1+\sum_{k=1}^\infty E^k\phi_k(x)\right]
\label{e09}
\end{equation}
and the derivative of $\psi(x)$ is given by
\begin{equation}
\psi'(x)=\frac{\psi_0'(x)}{\psi_0(x)}\psi(x)+\psi_0(x)\sum_{k=1}^\infty E^k
\phi_k'(x).
\label{e10}
\end{equation}
From (\ref{e10}), we immediately deduce the equation
\begin{equation}
\frac{\psi'(0)}{\psi(0)}=\psi_0'(0)+\sum_{k=1}^\infty E^k\phi_k'(0),
\label{e11}
\end{equation}
where on the right side of (\ref{e11}) we have used the relation $\psi(0)=
\psi_0(0)=1$, which holds in the perturbative regime (that is, inside the radius
of convergence of the power series in $E$). Accordingly, we define the function
$f(E)$ by
\begin{equation}
f(E)\equiv1-\frac{\psi'(0)}{\psi(0)\psi_0'(0)}.
\label{e12}
\end{equation}

We now wish to calculate the ground-state energy. The quantization condition
that determines the ground-state eigenvalue is simply that the slope of the
eigenfunction (\ref{e10}) vanish at $x=0$: $\psi'(0)=0$. If this condition
is satisfied, the ground-state eigenfunction is determined for all $x$,
negative as well as positive, by parity symmetry. Thus, an {\it implicit}
equation for the ground-state energy $E^{(0)}$ is $f\big(E^{(0)}\big)=1$.

The function $f(E)$ has a power series expansion of the general form
\begin{equation}
f(E)=\sum_{k=1}^\infty E^k a_k.
\label{e13}
\end{equation}
The first three terms in this series expansion for $f(E)$ are explicitly
\begin{eqnarray}
f(E)&=&-\frac{1}{\psi_0'(0)}\left[
E\int_0^\infty ds\,\psi_0^2(s)+E^2\int_0^\infty ds\,
\psi_0^2(s)\int_0^s dt\frac{1}{\psi_0^2(t)}\int_t^\infty du\,\psi_0^2(u)\right.
\nonumber\\
&&\!\!\!\!\!\!\!\!\!\!\!\!\!\!\!\!+\left. E^3\int_0^\infty ds\,\psi_0^2(s)
\int_0^s dt \frac{1}{\psi_0^2(t)}\int_t^\infty du\,\psi_0^2(u)\int_0^u dv\frac{
1}{\psi_0^2(v)}\int_v^\infty dw\,\psi_0^2(w)+\ldots~\right].
\label{e14}
\end{eqnarray}
It is clear from this expression that since $\psi_0'(0)$ is negative, the
coefficients $a_k$ are positive for all $k$. Because there are nested integrals,
the form of the equation $f(E)=1$ is reminiscent of a Rayleigh-Schr\"odinger
perturbation expansion \cite{R1} but the form of (\ref{e14}) is much simpler
because there is {\it no explicit reference to the potential} $V(x)$. The entire
dependence on the potential is contained in the function $\psi_0(x)$, which
satisfies the differential-equation boundary-value problem $\psi_0''(x)=V(x)
\psi_0(x)$, $\psi_0(0)=1$, $\psi_0(\infty)=0$.

The calculational scheme used here is the exact low-energy analog of the WKB
approximation, which is valid in the limit as $E\to\infty$. The WKB formula
for the $N$th eigenvalue (to all orders in WKB) is \cite{R2,R3,R4} 
\begin{equation}
\big(N+\half\big)\pi\sim\frac{1}{2i}\sum_{n=0}^\infty\oint_C dx\,S_{2n}(x),
\label{e15}
\end{equation}
where the contour $C$ encircles the two turning points in the positive sense.
The turning points are the two real solutions to $V(x)=E$, and $S_n(x)$ obeys
the recursion relation
\begin{eqnarray}
S_0(x)&=&-\sqrt{E-V(x)},\nonumber\\
S_1(x)&=& -\frac{S_0'(x)}{2S_0(x)},\nonumber\\
S_n(x)&=&
-\frac{1}{2S_0(x)}\left[S_{n-1}'(x)+\sum_{j=1}^{n-1}S_j(x)S_{n-j}(x)\right]
\quad(n\geq2).\nonumber
\end{eqnarray}

Note that the WKB series is normally thought of as a formal series in powers of
the ``small'' parameter $\hbar$. However, $\hbar$ is dimensional and cannot
actually be regarded as small. For that reason, we have not included it in the
series (\ref{e15}). The true small parameter in the WKB series is in fact $1/E$.
Indeed, evaluating the integrals in this WKB series typically produces a series
expansion in inverse (fractional) powers of $E$. For example, for the anharmonic
potential $V(x)=x^4$, the WKB series expansion reads
\begin{equation}
\big(N+\half\big)\pi\sim \sqrt{\pi}E^{3/4}\sum_{n=0}^\infty A_{2n}E^{-3n/2},
\label{e16}
\end{equation}
where the numerical coefficients $A_{2n}$ are given by
$$A_0=\frac{R}{3},\quad A_2=-\frac{1}{4\,R},\quad A_4=\frac{11\,R}{1536},\quad
A_6=\frac{4697}{30720\,R},\quad A_8=-\frac{390065\,R}{3670016},~\ldots$$
and $R=\Gamma\left(\frac{1}{4}\right)/\Gamma\left(\frac{3}{4}\right)$. We
emphasize that (\ref{e15}) and (\ref{e16}) are {\it implicit} representations
for $E$, and one must revert the series to find an explicit expression for $E_N$
as a series in powers of $1/N$. A significant difference between the two series
(\ref{e14}) and (\ref{e15}) is that while the WKB series (\ref{e15}) is
divergent, the series (\ref{e14}) is {\it convergent}. As we will see in
Sec.~\ref{s2}, the radius of convergence of (\ref{e14}) is finite and larger
than the ground-state energy $E^{(0)}$. In fact, the radius of convergence is
$E^{(1)}$, the energy of the first excited state; this is because $\psi(0)$ in
the denominator in (\ref{e12}) vanishes when $E=E^{(1)}$.

The series (\ref{e14}) determines more than just the ground-state energy. Since
$\psi(0)$ vanishes at all of the odd-parity eigenvalues and $\psi'(0)$ vanishes
at all of the even-parity eigenvalues, the function $f(E)-1$, as we can see in
(\ref{e12}), has simple poles at all the odd-parity eigenvalues and simple zeros
at all the even-parity eigenvalues. These distant poles and zeros are
inaccessible to the perturbation expansion used here. However, they become
accessible if the function $f(E)$ that the perturbation series represents is
analytically continued outside its radius of convergence. An approximate and
highly accurate analytic continuation is achieved by converting the truncated
series for $f(E)$ to a sequence of Pad\'e approximants. In Sec.~\ref{s6} we
construct explicitly the diagonal Pad\'e sequence for the illustrative
potentials considered in Sec.~\ref{s2} and obtain good numerical results.

This paper is organized as follows. In Sec.~\ref{s2} we consider potentials
of the form $V(x)=|x|^N$. To examine the accuracy of the procedure we consider
the special cases of the linear potential ($N=1$), the harmonic oscillator ($N=
2$), and the square-well potential ($N=\infty$). We also consider the nontrivial
case of the anharmonic oscillator $N=4$. In Sec.~\ref{s3} we show that the
numerical approximants obtained in Sec.~\ref{s2} are in a form that is ideally
suited for Shanks extrapolation and that a significant improvement in the
accuracy of the numerical results can be achieved by performing this
extrapolation procedure. Then, in Sec.~\ref{s4} we show that even greater
numerical accuracy can be achieved by using the approximate eigenfunction to
calculate the expectation value of the Hamiltonian. We extend our analysis to
the case of the $\cPT$-symmetric potentials and discuss the potential $V(x)
=ix^3$ in detail in Sec.~\ref{s5}. Next, in Sec.~\ref{s6} we calculate some
higher energy eigenvalues by the use of Pad\'e approximants. Section \ref{s7}
contains brief concluding remarks.

\section{Potentials of the form $V(x)=|x|^N$}
\label{s2}

In this section we apply the technique described in Sec.~\ref{s1} to potentials
of the form $V(x)=|x|^N$. For all such potentials $\psi_0(x)$ can be given in
closed form as an associated Bessel function:
$$\psi_0(x)=\frac{2(N+2)^{-1/(N+2)}}{\Gamma\left(\frac{1}{N+2}\right)}
x^{1/2}{\rm K}_{\frac{1}{2+N}}\left(\frac{x^{1+N/2}}{1+N/2}\right).$$

We can see from the structure of (\ref{e14}) that the coefficients $a_k$ in
(\ref{e12}) are all positive. Therefore, the graph of the function $f(E)$
passes through $0$ at $E=0$ and rises monotonically as $E$ increases. The value
of $E$ at which $f(E)$ passes through 1 is the ground-state energy $E^{(0)}$ of
$V(x)$. The graph of $f(E)$ continues to rise until $E$ reaches the first-energy
level $E^{(1)}$, the radius of convergence of the series. The function $f(E)$
becomes infinite at $E=E^{(1)}$.

To illustrate the calculation of $E^{(0)}$ we now consider in turn the cases $N=
\infty$, $N=1$, $N=2$, and $N=4$. For $N=\infty$, the coefficients $a_k$ are
rational numbers for all $k$, and for $N=1$ and $2$, the $a_n$ are known exactly
for all $k$ in terms of transcendental functions. For $N=4$, however, $a_k$ must
be calculated numerically as a $(2k-1)$-fold multiple integral.

\subsection{Square well potential $N=\infty$}
\label{ss21}

\begin{figure}[h!]
\begin{center}
\includegraphics[scale=0.60]{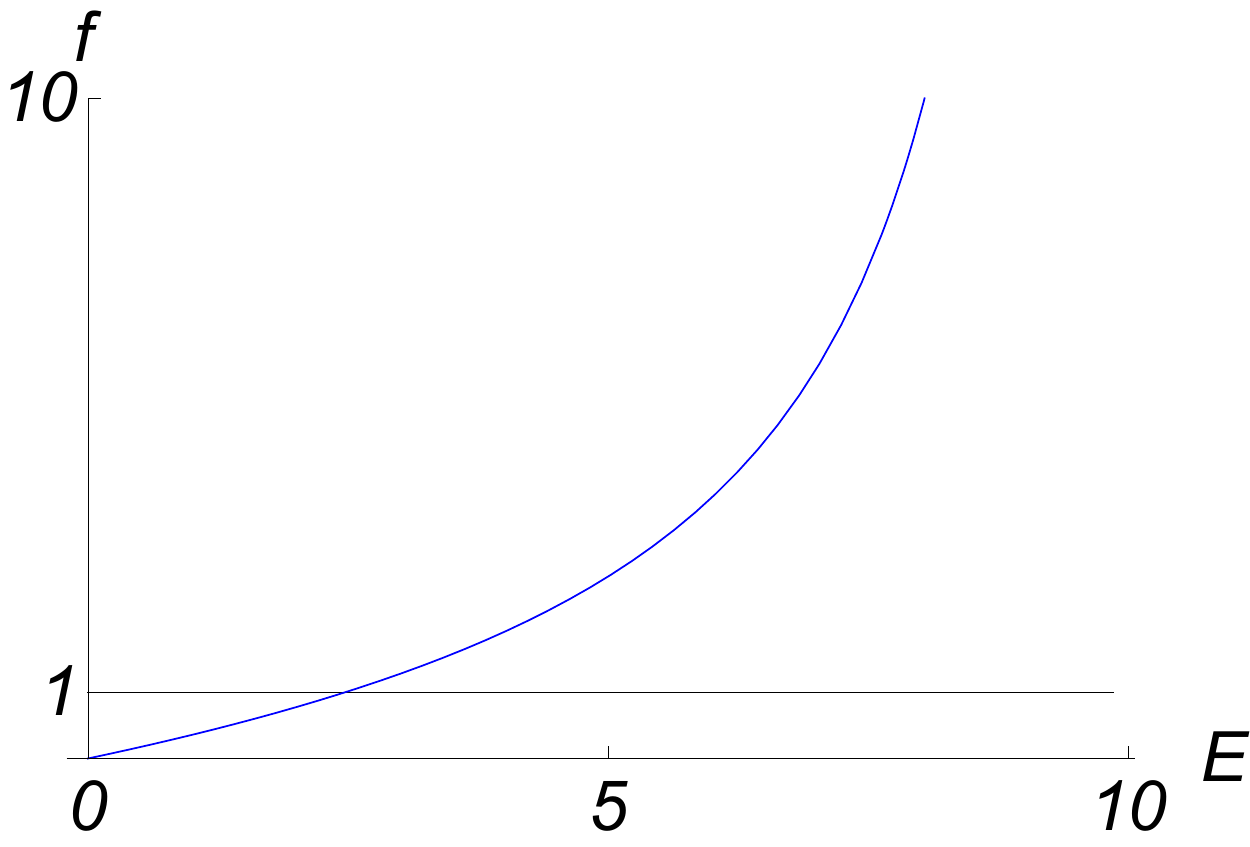}
\hspace{0.2cm}
\includegraphics[scale=0.60]{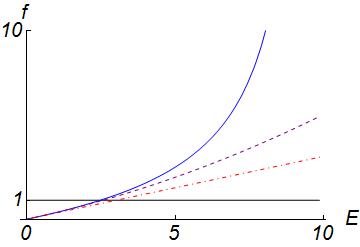}
\end{center}
\caption{Left panel: Plot of the function $f(E)$ in (\ref{e17}) for the case of
the square-well potential. The function $f(E)$ crosses 1 at $E=E^{(0)}=\pi^2/4=
2.46740$, the ground-state energy, and it becomes infinite at $E=E^{(1)}=\pi^2=
9.8696$, the radius of convergence of the series. Right panel: Same as the left
panel but with the first term (dashed-dotted line) and the first two terms
(dashed line) in the power-series expansion of $f(E)$ shown as well. One can see
from this graph that just a few terms in the series for $f(E)$ give an accurate
numerical approximation to $E^{(0)}$.}
\label{F1}
\end{figure}

For the special case of the square-well potential $V(x)=0$ ($|x|<1$), $V(x)=
\infty$ ($|x|\geq1$), we can find the function $f(E)$ in closed form by using
(\ref{e12}):
\begin{equation}
f(E)=1-\sqrt{E}\,{\rm cot}\sqrt{E}, 
\label{e17}
\end{equation}
which is valid for $E<E^{(1)}=\pi^2$. To obtain this result we have used 
$$\psi(x)=\frac{\sin\big[\sqrt{E}(1-x)\big]}{\sin\sqrt{E}}\quad{\rm and}\quad
\psi_0(x)=1-x$$
for $0\leq x\leq1$ and $E<\pi^2$. Figure~\ref{F1} shows that $f(E)$ vanishes at
$E=0$ and rises monotonically. It crosses 1 at $E=E^{(0)}=\pi^2/4$, the exact
value of the ground-state energy, and becomes infinite at $E=E^{(1)}=\pi^2$, the
radius of convergence of the Taylor series expansion of $f(E)$. The first two
partial sums of the Taylor series are also shown, and we can see graphically
that just a few terms in the Taylor series give an accurate approximation to
$E^{(0)}$. We can also see that the Taylor series converges monotonically upward
to $f(E)$.

We can expand $f(E)$ in (\ref{e17}) as a power series in $E$ and read off the
coefficients $a_k$:
\begin{equation}
f(E)=\frac{E}{3}+\frac{E^2}{45}+\frac{2E^3}{945}+\frac{E^4}{4725}+\frac{2E^5}
{93555}+\frac{1382E^6}{638512875}+\frac{4E^7}{18243225}+\cdots.
\label{e18}
\end{equation}
Truncating the series for $f(E)$ after $n$ terms and solving numerically for the
positive root $E_n$ of $f(E)-1$ gives a sequence of approximants $E_n$ to the
ground-state energy. The first six approximants $E_1,\,E_2,\,\ldots,E_6$ are
listed in Table~\ref{t1}. This table shows that $E_n$ converges geometrically to
the exact value $E^{(0)}=2.46740$ of the ground-state energy; it is shown in
Sec.~\ref{s3} that for large $n$ the difference between $E^{(0)}$ and $E_n$
approaches 0 like $4^{-n}$.

\begin{table}[h!]
\begin{tabular}{|c|c|c|}
\hline
{\bf Order} $n$&$E_n$&$E_n/E^{(0)}$\cr
\hline
1 & 3.0 & 1.21585\cr
2 & 2.56231 & 1.03846\cr
3 & 2.48906 & 1.00878\cr
4 & 2.47267 & 1.00214\cr
5 & 2.46871 & 1.00053\cr
6 & 2.46773 & 1.00013\cr
\hline
\end{tabular}
\caption{First six approximants $E_n$ ($n=1,\,2,\,\ldots,6$) to the exact value
of the ground-state energy $E^{(0)}=\pi^2/4=2.46740$. These approximants are
obtained by truncating the expansion of $f(E)-1$ after $n$ terms and finding the
positive root $E_n$ of the resulting polynomial. The convergence to $E^{(0)}$ is
geometric; the error in the $n$th approximant decays to zero like $4^{-n}$.}
\label{t1}
\end{table}

\subsection{Harmonic oscillator $N=2$}
\label{ss22}

\begin{figure}[h!]
\begin{center}
\includegraphics[scale=0.55]{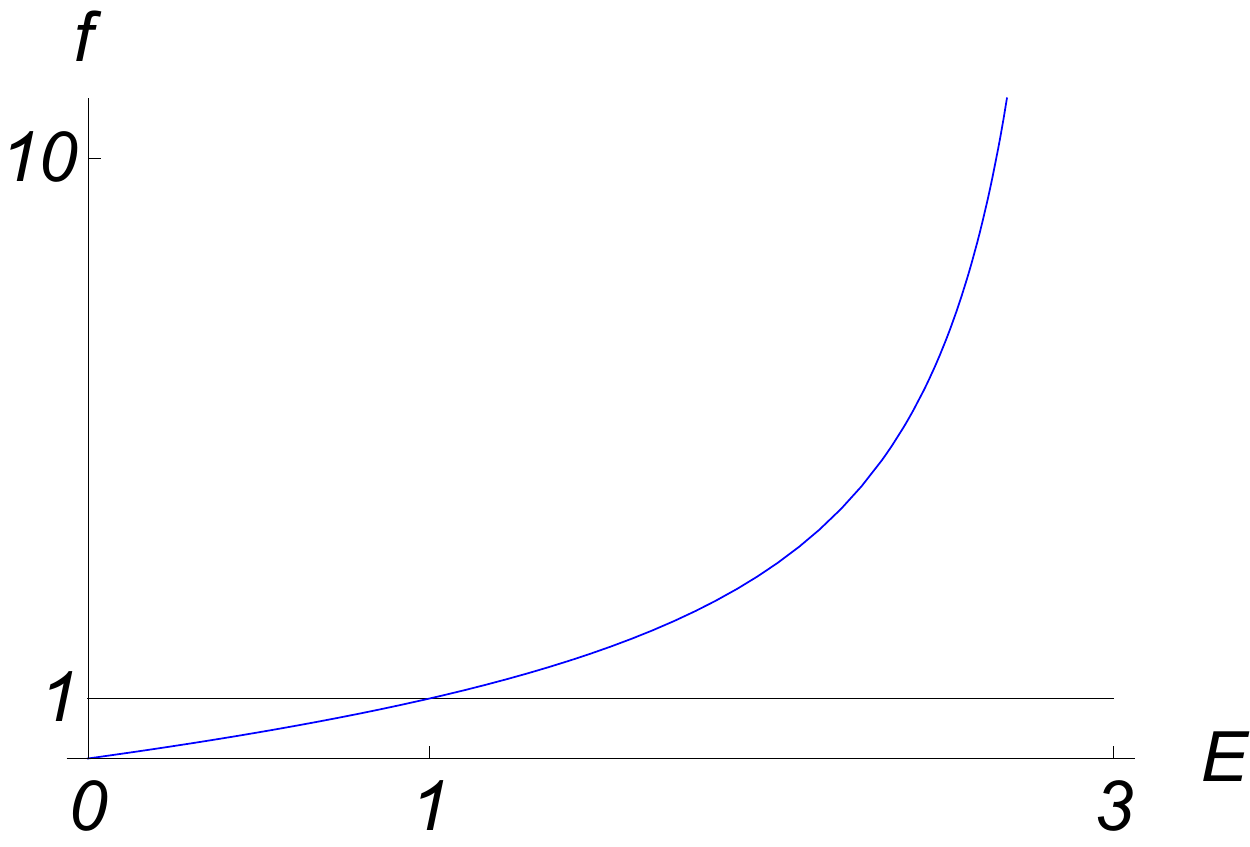}
\hspace{0.2cm}
\includegraphics[scale=0.55]{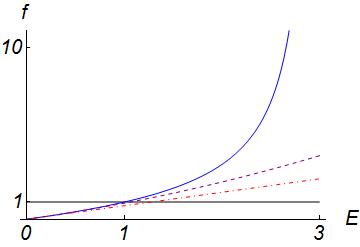}
\end{center}
\caption{Left panel: Plot of $f(E)$ for the potential $V(x)=x^2$ of the quantum
harmonic-oscillator. Note that $f(E)=1$ at $E=E^{(0)}=1$ and becomes
infinite when $E=E^{(1)}=3$. Right panel: Same as the left panel but with the
first two partial sums of the power series expansion for $f(E)$ shown as
well. Note that the partial sums converge monotonically upward to $f(E)$.}
\label{F2}
\end{figure}

For the harmonic-oscillator potential $V(x)=x^2$, the function $f(E)$ is
\begin{equation}
f(E)=1-\frac{\Gamma(1/4)\Gamma(3/4-E/4)}{\Gamma(3/4)\Gamma(1/4-E/4)}.
\label{e19}
\end{equation}
We obtain this result from 
$$\psi(x)=\frac{{\rm D}_{(E-1)/2}\big(x\sqrt{2}\big)}{{\rm D}_{(E-1)/2}(0)},
\quad\psi_0(x)=\frac{{\rm D}_{-1/2}\big(x\sqrt{2}\big)}{{\rm D}_{-1/2}(0)}
\quad\big(E<E^{(1)}=3\big),$$
where ${\rm D}_\nu(z)$ is the parabolic cylinder function \cite{R5}. The
power-series expansion of this function gives the coefficients $a_k$:
\begin{equation}
f(E)=0.78530E+0.14956E^2+0.04403E^3+0.01409E^4+0.00463E^5+0.00153E^6+\cdots,
\label{e20}
\end{equation}
which come from evaluating polylogarithms. As Fig.~\ref{F2} illustrates, $f(E)$
vanishes at $E=0$, crosses $1$ at the ground-state energy $E^{(0)}=1$ and
becomes infinite at $E^{(1)}=3$. Table \ref{t2} lists the first six approximants
$E_n$ to $E^{(0)}$. In Sec.~\ref{s3} it is shown that the error vanishes
geometrically for large $n$ like $3^{-n}$.

\begin{table}[h!]
\begin{tabular}{|c|c|}
\hline
{\bf Order} $n$&$E_n$\cr
\hline
1 & 1.27324 \cr
2 & 1.05949 \cr
3 & 1.01721 \cr
4 & 1.00543 \cr
5 & 1.00177 \cr
6 & 1.00059 \cr
\hline
\end{tabular}
\caption{First six approximants $E_n$ to the ground-state energy $E^{(0)}=1$ of
the harmonic oscillator. The error is of order $3^{-n}$.}
\label{t2}
\end{table}

\subsection{Linear potential $N=1$}
\label{ss23}

For the linear potential $V(x)=|x|$, the function $f(E)$ is
\begin{equation}
f(E)=1-\frac{{\rm Ai}(0){\rm Ai}'(-E)}{{\rm Ai}'(0){\rm Ai}(-E)},
\label{e21}
\end{equation}
which is derived by substituting $\psi(x)={\rm Ai}(x-E)/{\rm Ai}(-E)$ into
(\ref{e12}). The Taylor expansion of this function gives the coefficients $a_k$:
\begin{equation}
f(E)=0.72901E+0.15440E^2+0.05411E^3+0.02131E^4+0.00876E^5+0.00368E^6+\cdots,
\label{e22}
\end{equation}
which series converges if $E<E^{(1)}$. The function $f(E)$ is plotted in
Fig.~\ref{F2}; $f(E)$ crosses $1$ at the ground-state energy $E^{(0)}=
1.01879297$ and becomes infinite at $E^{(1)}=2.3381075$. The first six
approximants $E_n$ to the ground-state energy $E^{(0)}$ are listed in
Table~\ref{t3}. It is shown in Sec.~\ref{s3} that the error in the $n$th
approximant vanishes for large $n$ like $2.295^{-n}$.

\begin{figure}[h!]
\begin{center}
\includegraphics[scale=0.55]{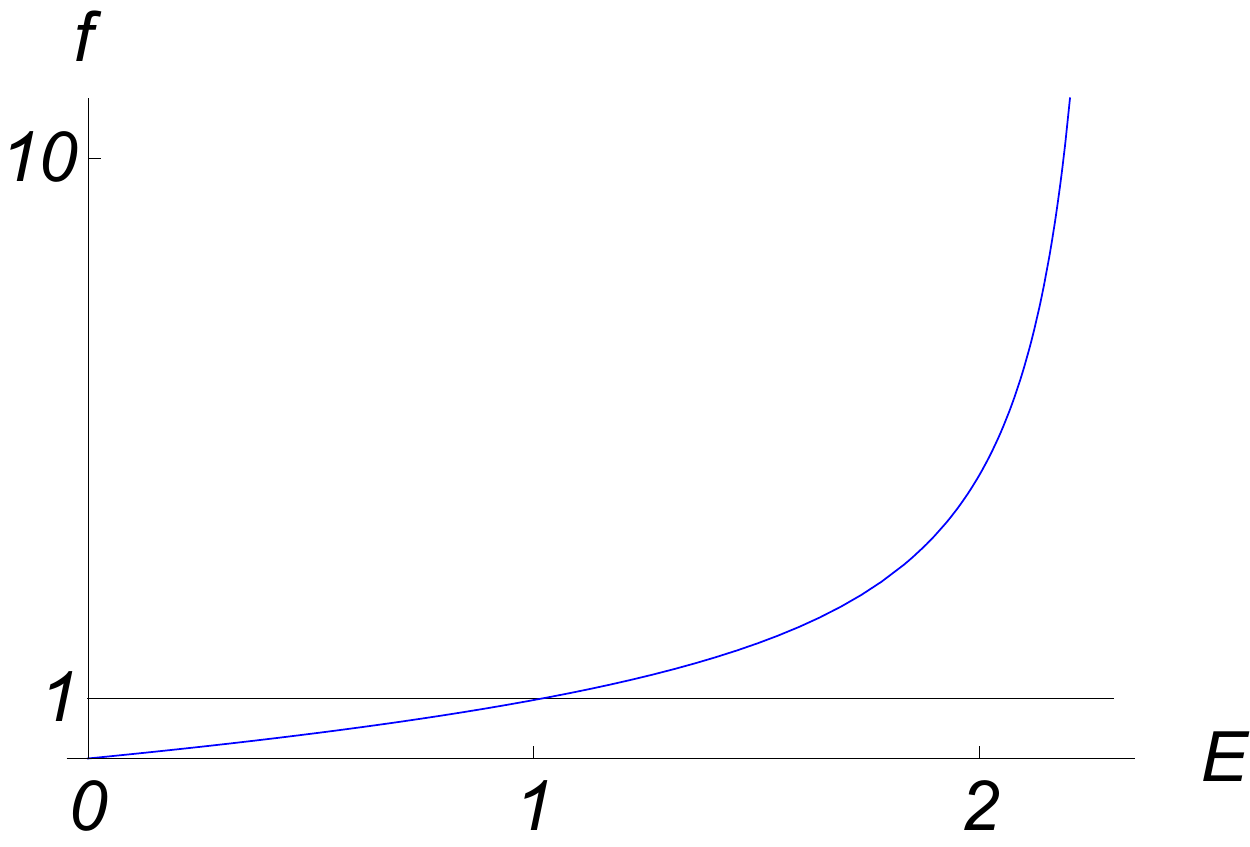}
\hspace{0.2cm}
\includegraphics[scale=0.55]{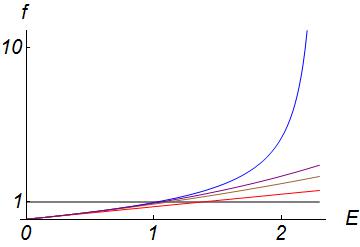}
\end{center}
\caption{Left panel: Plot of $f(E)$ in (\ref{e22}) for the linear potential $V(x
)=|x|$. Note that $f(E)$ passes through $1$ at the ground-state energy $E=^{(0)}
1.01879297$ and becomes infinite at $E^{(1)}=2.3381075$. Right panel: Same as
left panel but with the first three partial sums in the power-series expansion
for $f(E)$ shown as well. Note that the convergence to $f(E)$ is monotone upward
and that just a small number of terms in the series gives an accurate value of
$E^{(0)}$.}
\label{F3}
\end{figure}

\begin{table}[h!]
\begin{tabular}{|c|c|c|}
\hline
{\bf Order} $n$&$E_n$&$E_n/E^{(0)}$\cr
\hline
1 & 1.37172 & 1.34642\cr
2 & 1.11052 & 1.09003\cr
3 & 1.05136 & 1.03197\cr
4 & 1.03168 & 1.01265\cr
5 & 1.02415 & 1.00525\cr
6 & 1.02107 & 1.00223\cr
\hline
\end{tabular}
\caption{First six approximants $E_n$ to the ground-state energy of the linear
oscillator potential $V(x)=|x|$. The exact value of the ground-state energy is
$E^{(0)}=1.01879$. The error in the $n$th approximant is of order $2.295^{-n}$.}
\label{t3}
\end{table}

\subsection{Quartic potential $N=4$}
\label{ss24}

For the quartic potential it is not easy to calculate many terms in the Taylor
series expansion of $f(E)$ because it requires the numerical evaluation of
multiple integrals. However, the first three terms in this series are
\begin{equation}
f(E)=0.763303 E+0.125262 E^2+0.030303 E^3+\cdots.
\label{e23}
\end{equation}
In Table \ref{t4} we give the results of calculating the successive
zeros of this series.

\begin{table}[h!]
\begin{tabular}{|c|c|c|}
\hline
{\bf Order} $n$&$E_n$&$E_n/E^{(0)}$\cr
\hline
1 & 1.31010 & 1.23552\cr
2 & 1.10846 & 1.04536\cr
3 & 1.07240 & 1.01136\cr
\hline
\end{tabular}
\caption{First three approximants to the ground-state energy $1.060362$ of the
anharmonic oscillator potential $V(x)=x^4$. These approximants are obtained by
truncating the series (\ref{e23}).}
\label{t4}
\end{table}

\section{Shanks extrapolation}
\label{s3}

The Shanks transformation \cite{R6} is a technique for finding the limit $L$ of
a sequence $\{A_n\}$ as $n\to\infty$. This technique relies on the assumption
that the $n$th term in the sequence has the asymptotic form
\begin{equation}
A_n\sim L+cr^n\quad(n\to\infty),
\label{e24}
\end{equation}
where $c$ and $r$ are constants and $|r|<1$. If the asymptotic approximation
(\ref{e24}) is accurate, then a simultaneous solution to the three equations
\begin{equation}
A_{n+1}=L+cr^{n+1},\quad A_n=L+cr^n,\quad A_{n-1}=L+cr^{n-1}
\label{e25}
\end{equation}
gives an accurate value for the limit $L$:
\begin{equation}
L=\frac{A_{n+1}A_{n-1}-A_n^2}{A_{n+1}+A_{n-1}-2A_n}.
\label{e26}
\end{equation}

Of course, a sequence typically has a more complicated form than the simple
three-parameter model in (\ref{e24}), and therefore the value of $L$ in 
(\ref{e26}) is only approximate. Indeed, this calculation produces a value of
$L$ that depends on $n$. However, if the three-parameter model (\ref{e24}) is
accurate, then the formula in (\ref{e26}) produces a new sequence
\begin{equation}
\big\{\cS\big(A_n\big)\big\}\equiv\frac{A_{n+1}A_{n-1}-A_n^2}{A_{n+1}+A_{n-1}
-2A_n}
\label{e27}
\end{equation}
that typically converges to the limit $L$ faster than the sequence $\{A_n\}$ as
$n\to\infty$. The new sequence is called the {\it Shanks transform} of the
original sequence $\{A_n\}$.

The ground-state energy $E^{(0)}$ is the positive solution to the equation $1=
f(E)$, where $f(E)$ is given in (\ref{e12}). Let $E_n$ be the
positive root of the polynomial equation $1=f_n(E)$, where $f_n(E)$ is the $n$th
partial sum of $f(E)$:
\begin{equation}
1=\sum_{k=1}^n a_k\big(E_n\big)^k.
\label{e28}
\end{equation}
Also, $E_{n+1}$ is the positive root of polynomial equation
\begin{equation}
1=\sum_{k=1}^{n+1}a_k\big(E_{n+1}\big)^k.
\label{e29}
\end{equation}
If we subtract (\ref{e28}) from (\ref{e29}), we obtain the equation
\begin{equation}
0=\sum_{k=1}^n a_k\left[\big(E_{n+1}\big)^k-\big(E_n\big)^k\right]+
a_{n+1}\big(E_{n+1}\big)^{n+1}.
\label{e30}
\end{equation}

Our numerical studies of $f(E)$, as described in Sec.~\ref{s2}, show that for
large $n$, $E_n$ is approximated very well by the simple Shanks formula
\begin{equation}
E_n\sim E^{(0)}+cr^n.
\label{e31}
\end{equation}
Therefore, we can make the approximations
\begin{eqnarray}
\big(E_n\big)^k&\sim&\big(E^{(0)}\big)^k\big(1+kcr^n/E^{(0)}\big),\nonumber\\
\big(E_{n+1}\big)^k&\sim&\big(E^{(0)}\big)^k\big(1+kcr^{n+1}/E^{(0)}\big).
\label{e32}
\end{eqnarray}
Furthermore, the power series representation for $f(E)$ in (\ref{e12}) has a
nonzero radius of convergence, which we denote here by $R$. Thus, for large $n$,
we can approximate the $n$th coefficient $a_n$ in the series by $KR^{-n}$, where
$K$ is a constant. (In addition to this geometric dependence there may also be
an algebraic dependence on $n$, but such a dependence does not affect this
argument.) Thus, we can approximate the last term in (\ref{e30}) by
\begin{equation}
a_{n+1}\big(E_{n+1}\big)^{n+1}\sim K\big(E^{(0)}\big)^{n+1}/R^{n+1}.
\label{e33}
\end{equation}

These approximations simplify the formula in (\ref{e30}) to
\begin{equation}
(1-r)c r^nf'\big(E^{(0)}\big)\sim K\big(E^{(0)}\big)^{n+1}/R^{n+1}.
\label{e34}
\end{equation}
Thus, in the limit as $n\to\infty$ we obtain equations for $r$ and $c$:
\begin{equation}
r=E^{(0)}/R,\quad c=\frac{KE^{(0)}}{\big[R-E^{(0)}\big]f'\big[E^{(0)}\big]}.
\label{e35}
\end{equation}

Let us apply this analysis in turn to the square-well, harmonic-oscillator,
linear, and quartic potentials. For the square-well potential $f(E)$ is given in
(\ref{e17}), and we can see from this formula that $E^{(0)}=\pi^2/4$ and that $R=
E^{(1)}=\pi^2$. Thus, (\ref{e35}) implies that $r=1/4$. This explains the
observed rate of convergence of the approximants $E_n$ in Table~\ref{t1}. If we
now compute the Shanks transform $\cS\big[E_n/E^{(0)}\big]$ of the entries in the
third column in Table~\ref{t1}, we obtain the new improved sequence of
approximants $1.00281$, $1.00022$, $1.00002$, $1.00000$, which is a dramatic
improvement in accuracy. [Note that the six entries in Table~\ref{t1} can only
give rise to four terms in the Shanks transformed sequence because of the
structure of (\ref{e26}).]

For the harmonic-oscillator potential $f(E)$ is given in (\ref{e19}). Thus, $E^{
(0)}=1$ and $R=E^{(1)}=3$. Thus, from (\ref{e35}) we see that $r=1/3$. This
explains the observed rate of convergence of the approximants $E_n$ in
Table~\ref{t2}. We compute the Shanks transform of the entries in the second
column in Table~\ref{t2} and we obtain the new improved sequence of approximants
$1.00678$, $1.00088$, $1.00012$, $1.00002$, which again is a dramatic improvement
in accuracy.

Next, we consider the linear potential; $f(E)$ given in (\ref{e21}). The value of
$r=E^{(0)}/E^{(1)}$ is $1/2.29459$, which explains the rate of convergence of the
approximants in Table~\ref{t3}. The Shanks transform of the entries in the third
column in produces the new and more accurate sequence of approximants $1.01497$,
$1.00301$, $1.00066$, $1.00014$.

Finally, we construct the Shanks transform of the three entries in the third
column in Table~\ref{t4} and obtain $1.00396$. This is an improvement in accuracy
by a factor of three.

\section{Expectation value of $H$}
\label{s4}

Let us denote by $\Psi_n(x)$ the truncation of the series (\ref{e09}) for $\psi
(x)$ at order $n$,
\begin{equation}
\Psi_n(x)=\psi_0(x)\left[1+\sum_{k=1}^n E^k\phi_k(x)\right]
\label{e36}
\end{equation}
in which we replace $E$ by $E_n$ so that $\Psi_n(x)$ satisfies the boundary
condition $\Psi_n'(0)=0$. The expectation value $\langle H\rangle_n$, of the
operator $H$ in the state $\Psi_n(x)$ is given by
\begin{equation}
\langle H\rangle_n=\frac{1}{\cal N}\int_0^\infty dx\,\Psi_n(x)\left[-\Psi_n''(x)
+x^N\Psi_n(x)\right],
\label{e37}
\end{equation}
where ${\cal N}=\int_0^\infty dx\,\big[\Psi_n(x)\big]^2$. Note that we have
taken the integration ranges as $(0,\infty)$ rather than $(-\infty,\infty)$
because $\Psi_n(x)$ is constructed as an even function when $E=E_n$.

Using (\ref{e03}) and the recursion relations (\ref{e06}) for the $\phi_k$ we
readily find that
\begin{equation}
H\Psi_n\equiv-\Psi_n''+x^N\Psi_n=E\Psi_{n-1}.
\label{e38}
\end{equation}
This result reveals the extent to which $\Psi_n$ fails to satisfy the
Schr\"odinger equation, for which the right side of this equation would be $E
\Psi_n$. Using this equation, we get
\begin{equation}
\langle H\rangle_n=E_n\frac{\int_0^\infty dx\,\Psi_n(x)\Psi_{n-1}(x)}
{\int_0^\infty dx\,\big[\Psi_n(x)\big]^2}.
\label{e39}
\end{equation}
Because all the terms in the expansion of $\psi(x)$ are positive, $\Psi_{n-1}(x)
<\Psi_n(x)$. Thus, we see that $\langle H\rangle_n<E_n$. But, the expectation
value of $H$ in any approximate eigenfunction must satisfy the inequality
$\langle H\rangle_n>E_{\rm exact}$. Thus,
\begin{equation}
E_n>\langle H\rangle_n>E_{\rm exact}.
\label{e40}
\end{equation}
So, taking the expectation value of $H$ in the state $\Psi_n$ is guaranteed to
produce a more accurate approximation to $E^{(0)}$ than $E_n$. When the $\phi_k$
must be calculated as multiple integrals, the maximal dimension of the integrals
involved in calculating $\langle H\rangle_n$ is $2n+1$, to be compared with $2n-
1$ for $E_n$. Thus, the maximal dimension, and hence the computational effort,
is the same for $\langle H \rangle_n$ and $E_{n+1}$. However, we shall see that
in every case $\langle H\rangle_n$ is much more accurate than $E_n$.

Let us now consider in turn the potentials studied in Sec.~\ref{s2} and compare
$\langle H\rangle_n$ with the results given in Tables \ref{t1} - \ref{t4}. For
the square well, case A, we have the results that
\begin{eqnarray*}
\langle H \rangle_1/E_{\rm exact}&=& 1.001292,\cr
\langle H \rangle_2/E_{\rm exact}&=& 1.000061,\cr
\langle H \rangle_3/E_{\rm exact}&=& 1.000003.
\end{eqnarray*}
For the quantum harmonic oscillator, case B, the results are
\begin{eqnarray*}
\langle H \rangle_1 &=& 1.003921,\cr
\langle H \rangle_2 &=& 1.000343,\cr
\langle H \rangle_3 &=& 1.000035.
\end{eqnarray*}
For the linear potential, case C, we have
\begin{eqnarray*}
\langle H \rangle_1/E_{\rm exact}&=& 1.009813,\cr
\langle H \rangle_2/E_{\rm exact}&=& 1.001427,\cr
\langle H \rangle_3/E_{\rm exact}&=& 1.019041.
\end{eqnarray*}
Finally, for the anharmonic oscillator, case C, we have
\begin{eqnarray*}
\langle H \rangle_1/E_{\rm exact}&=& 1.00202,\cr
\langle H \rangle_2/E_{\rm exact}&=& 1.00012.
\end{eqnarray*}
These results become more accurate for larger values of the power $N$ of the
potential. This may be because the wave functions fall off more rapidly with
$x$ as $N$ increases, so the expectation values are less sensitive to the
difference in shape between $\Psi_n$ and $\Psi_{n-1}$.

\section{$\cPT$-Symmetric Potential $ix^3$}
\label{s5}

Until now we have dealt with real symmetric potentials, $V(-x)=V(x)$, and have
exploited the symmetry of the ground-state eigenfunction. The approximate
solutions have therefore been required to satisfy the condition $\Psi_n'(0)=0$.
However, our method readily extends to complex ${\cal PT}$-symmetric potentials
satisfying $V(-x)=V^*(x)$. In particular, it has been shown in a number of
papers \cite{R7,R8,R9} that the potentials $V=-(ix)^N$ for $N\ge2$ have a
completely real energy spectrum. When $N\geq2$, the ${\cal PT}$ symmetry is {\it
unbroken}, that is, the phases of the eigenfunctions can be chosen so that the
eigenfunctions are ${\cal PT}$ symmetric, $\psi^*(-z)=\psi(z)$. Here, we have
written the argument as $z$ because the ${\cal PT}$-symmetric eigenvalue problem
can, and for $N\ge4$ must, be posed on a contour in the complex-$z$ plane lying
within an appropriate Stokes wedge \cite{R7}. We choose the contour to pass
through the origin, and the corresponding condition on the truncated wave
functions $\Psi_n(z)$ in (\ref{e36}) is
\begin{eqnarray}
{\rm Re}\left(\frac{1}{\psi(z)}\frac{d\psi(z)}{dz}\right)_{z=0}=0.
\label{e41}
\end{eqnarray}
It is convenient to formulate the problem in the right-half plane on the Stokes
line $z=\lambda x$, where $\lambda=e^{-i\theta }$ and $\theta=\pi/(2N+4)$. Then
for $N=3$ the Schr\"odinger equation on the Stokes line reads
\begin{eqnarray}
\left(-\frac{d^2}{dx^2}+ x^3\right)\psi(x)=\lambda^2 E\psi(x).
\label{e42}
\end{eqnarray}
The coefficients in the energy power-series expansion of $\psi(x)$ are the same
as those for the potential $V=|x|^3$. The only difference between the $ix^3$ and
the $|x|^3$ potentials is that now the expansion is in powers of the complex
quantity $\lambda^2E$ instead of $E$, where in this case $\theta=\pi/10$.

In terms of $x$ the eigenvalue condition at $z=0$ becomes
\begin{equation}
{\rm Re}\left(\frac{1}{\lambda}\frac{1}{\psi(x)}\frac{d\psi(x)}{dx}\right)_{x=0}
=0.
\label{e43}
\end{equation}
Thus,
\begin{eqnarray}
\cos(\theta)\,\frac{\psi'_0(0)}{\psi_0(0)}+\cos(\theta)\,E\phi'_1(0)+\cos(3
\theta)\,E^2\phi'_2(0)+\dots=0.
\label{e44}
\end{eqnarray}

The principal difference from the Hermitian case is that the coefficient $a_k$
in (\ref{e13}) is multiplied by the factor $\cos{(2k-1)\theta}$, which is not
positive definite. As a result, some of the coefficients are negative, and other
coefficients vanish. This means that convergence of $E_n$ to the exact
ground-state energy $E^{(0)}$ is no longer monotonic. These features are
exemplified by the following numerical results:
\begin{eqnarray}
E_1/E^{(0)} &=& 1.10366,\nonumber\\
E_2/E^{(0)} &=& 0.98258,\nonumber\\
E_3/E^{(0)} &=& E_2/E^{(0)}.
\label{e45}
\end{eqnarray}
The first-order result happens to be the same as for $V=|x|^3$ because $\cos
(-\theta)=\cos{\theta}$. The second-order result is {\it less} than $E^{(0)}$,
rather than approaching the exact value from above, and the approximant is
unchanged in third order because $\cos(5\theta)=0$.

The reason for choosing the contour as we did is that the integrals over $x$
that are used to construct $\phi_k(x)$ and $\phi_k'(x)$ remain real, and in fact
equal those for the Hermitian potential $|x|^3$, thus making the evaluation of
the coefficients no more difficult than in the Hermitian case. The difference is
that in calculating the expectation value of $H$ according to (\ref{e39}), the
truncated wave functions $\Psi_n$ and $\Psi_{n-1}$ are complex because of the
replacement of $E_n$ by $\lambda^2E_n$ in the expansion (\ref{e02}). However, by
taking the real parts the integrals can be decomposed into a number of real
integrals over the $\phi_k(x)$. The results are
\begin{eqnarray}
\langle H\rangle_1/E^{(0)} &=& 0.984,\nonumber\\
\langle H\rangle_2/E^{(0)} &=& 0.997.
\label{e46}
\end{eqnarray}
Again, both $\langle H\rangle_1$ and $\langle H\rangle_2$ are less than the true
value $E^{(0)}$ but closer than $E_2$ and $E_3$, respectively. Recall that
$\langle H\rangle_n$ and $E_{n+1}$ require about the same calculational effort.

In summary, we can see that while the results do converge to the ground-state
energy, the convergence is not as rapid as for the corresponding Hermitian
potential $V=|x|^N$. This is not surprising considering the very similar
variational results in Ref.~\cite{R10}, where high-dimensional matrices were
diagonalized to obtain the numerical approximations to the eigenvalues.

\section{Pad\'e calculation of higher energy levels}
\label{s6}

As explained in Sec.~\ref{s1}, one can find the approximate poles and zeros
of $f(E)-1$ by converting (\ref{e14}) to the diagonal sequence of Pad\'e 
approximants. The poles give approximations to the odd-parity eigenvalues,
while the zeros give approximations to the even-parity eigenvalues. In Tables
\ref{t5}-\ref{t7} below we list the first four eigenvalues obtained from the
Pad\'e approximants $P_1^1$, $P_2^2$, $P_3^3$, and $P_4^4$ for the
potentials considered in Subsecs.~\ref{ss21}-\ref{ss23}.

\begin{table}[h!]
\begin{tabular}{|c|c|c|c|c|c|}
\hline
{\bf Energy} & $P_1^1$ & $P_2^2$ & $P_3^3$ & $P_4^4$ & {\bf Exact} \cr
\hline
$E^{(0)}$ & $2.50000$ & $2.46744$ & $2.46740$ & $2.46740$ & $2.46740$\cr
$E^{(1)}$ & ---       & $9.94122$ & $9.86993$ & $9.86960$ & $9.86960$\cr
$E^{(2)}$ & ---       & ---       & $22.29341$ & $22.20737$ & $22.20661$\cr
$E^{(3)}$ & ---       & ---       & ---        & $39.56379$ & $39.47842$\cr
\hline
\end{tabular}
\caption{Energies obtained from the first four diagonal Pad\'e approximants for
the square-well potential $V(x)=0$ ($|x|\leq1)$, $V(x)=\infty$ ($|x|>1$).}
\label{t5}
\end{table}

\begin{table}[h!]
\begin{tabular}{|c|c|c|c|c|c|}
\hline
{\bf Energy} & $P_1^1$ & $P_2^2$ & $P_3^3$ & $P_4^4$ & {\bf Exact} \cr
\hline
$E^{(0)}$ & $1.02478$ & $1.00013$ & $1.00000$ & $1.00000$ & $1$\cr
$E^{(1)}$ & ---       & $3.08260$ & $3.00237$ & $3.00003$ & $3$\cr
$E^{(2)}$ & ---       & ---       & $5.12647$ & $5.00701$ & $5$\cr
$E^{(3)}$ & ---       & ---       & ---       & $7.16012$ & $7$\cr
\hline
\end{tabular}
\caption{Energies obtained from the first four diagonal Pad\'e approximants for
the harmonic oscillator potential $V(x)=x^2$.}
\label{t6}
\end{table}

\begin{table}[h!]
\begin{tabular}{|c|c|c|c|c|c|}
\hline
{\bf Energy} & $P_1^1$ & $P_2^2$ & $P_3^3$ & $P_4^4$ & {\bf Exact} \cr
\hline
$E^{(0)}$ & $1.06291$ & $1.01948$ & $1.01880$ & $1.01879$ & $1.01879$\cr
$E^{(1)}$ & ---       & $2.48513$ & $2.34902$ & $2.33863$ & $2.33811$\cr
$E^{(2)}$ & ---        & ---      & $3.44920$ & $3.27292$ & $3.24820$\cr
$E^{(3)}$ & ---        & ---      & ---       & $4.35282$ & $4.08795$\cr
\hline
\end{tabular}
\caption{Energies obtained from the first four diagonal Pad\'e approximants for
the linear potential $V(x)=|x|$.}
\label{t7}
\end{table}

For the anharmonic oscillator potential $V(x)=x^4$ we have only calculated the
series for $f(E)$ in (\ref{e23}) to third order in powers of $E$. If we use the
diagonal Pad\'e approximant $P_1^1$, we obtain the estimate $E^{(0)}=1.07827$,
which is accurate to about 2\% and is a slight improvement on the second entry
in Table~\ref{t4}. However, the Pad\'e approximant $P_1^2$ gives $E^{(0)}=
1.06137$, a significant improvement in numerical accuracy (one part in a
thousand) compared with the results in Table~\ref{t4}. This Pad\'e approximant
also gives the value $4.13364$ for the first excited state, which differs from
the exact value $3.79967$ by about 9\%.

\section{BRIEF COMMENTS}
\label{s7}

In this paper we have shown how to construct a perturbative solution to the
time-independent Schr\"odinger equation $-\psi''(x)+V(x)\psi(x)=E\psi(x)$ [$V(x)$
even] as a formal series in powers of the energy $E$ itself. We have then used
this expansion to obtain remarkably accurate numerical approximations to the
ground-state energy, and also to the higher energy levels by the use of various
numerical methods such as the Shanks transformation and Pad\'e approximation. The
surprise is that even though the energy levels of the quantum theory are not
small compared with 1, the perturbation expansion is {\sl convergent} if $E<E^{
(1)}$, the first energy level. Furthermore, the perturbation expansion that we
have constructed can also be applied to non-Hermitian $\cPT$-symmetric
potentials. Our general approach has been to solve the unperturbed problem $-
\psi_0(''x)+V(x)\psi_0(x)=0$ for $x\ge0$ and to use $\psi_0(x)$ as the building
block for constructing $\psi(x)$ for $x\ge0$ as a perturbative expansion in $E$.
The approximate eigenvalues are then obtained by the condition that $\psi(x)$ can
be extended by parity (or $\cPT$-symmetry) to negative $x$. In the future, we
plan to extend the techniques developed here to quantum theories in
higher-dimensional space.

Finally, in the present paper we have limited ourselves to providing an
elementary, accurate, and general recipe for calculating the low-lying
eigenvalues of symmetric or $\cPT$-symmetric potentials using simple analytical
and numerical tools. The recipe essentially provides a method of calculating the
truncated Weierstrass products for $\psi'(0)$ and $\psi(0)$. Hence it may be
interesting to explore further the connection of the method to such topics as
spectral zeta functions, functional determinants and infinite products of
eigenvalues [11].

\begin{acknowledgments}
CMB was supported by a grant from the U.S.~Department of Energy.
\end{acknowledgments}

\end{document}